\documentclass[reprint,amsmath,amssymb,aps,]{revtex4-1}

\usepackage{amsmath}
\usepackage{graphicx}
\usepackage{dcolumn}
\usepackage{bm}
\usepackage[normalem]{ulem}
\usepackage[colorlinks=true,linkcolor=blue,urlcolor=blue,citecolor=blue]{hyperref}
\usepackage{soul}

\begin{document}

\title{Evidence for degenerate mirrorless lasing in alkali metal vapor: forward beam magneto-optical experiment}

\author{Aram Papoyan$^{1}$}
\author{Svetlana Shmavonyan$^{1}$}
\author{Aleksandr Khanbekyan$^{1}$}
\author{Hrayr Azizbekyan$^{1}$}
\author{Marina Movsisyan$^{1}$}
\author{Guzhi Bao$^{2,3}$}
\author{Dimitra Kanta$^{2}$}
\author{Arne Wickenbrock$^{2}$}
\author{Dmitry Budker$^{2,4,5}$}
\affiliation{
$^{1}$ Institute for Physical Research, National Academy of Sciences of Armenia, 0203 Ashtarak-2, Armenia\\
$^{2}$ Helmholtz Institute Mainz, Johannes Gutenberg University, 55099 Mainz, Germany\\
$^{3}$ Department of Physics, East China Normal University, Shanghai 200062, P. R. China. \\
$^{4}$ Department of Physics, University of California, Berkeley, CA 94720-7300, USA\\
$^{5}$ Nuclear Science Division, Lawrence Berkeley National Laboratory, Berkeley, CA 94720, USA}
\begin{abstract}
We report an experimental observation of degenerate mirrorless lasing in forward direction under excitation of a dilute atomic Rb vapor with a single linearly polarized cw laser light resonant with cycling $F_e > F_g$ atomic D$_2$ transitions. Light polarized orthogonally to the laser light is generated for the input light intensity exceeding a threshold value of $\sim$ 3 mW/cm$^2$. Application of a transverse magnetic field directed along the input light polarization reveals a sharp $\sim$ 20 mG wide magnetic resonance centered at $B$ = 0. Increasing the incident light intensity from 3 to 300 mW/cm$^2$, the generated light undergoes rapid amplitude increase followed by a decline and resonance broadening. Such nonlinear behavior of the observed magnetic resonance is attributed to the population inversion on optical transitions between magnetic sublevels established under linearly polarized excitation. We present observations that indicate that a combination of nonlinear-optical effects occurs in this system, including degenerate mirrorless lasing and four-wave mixing.
\end{abstract}
\maketitle
\section{Introduction}
Nonlinear magneto-optical phenomena in alkali-metal vapors such as nonlinear Faraday, Voigt and Hanle effects resulting in resonant dispersive or absorptive features in the polarization of transmitted or scattered light in the presence of a magnetic field have been extensively studied both experimentally and theoretically (see \cite{budker1} and references therein). Recent interest towards these processes is stimulated by the possibility of substantial enhancement of nonlinear magneto-optical effects exploiting long-lived ground-state coherences, which is particularly important for development of sensitive optical magnetometers \cite{budker2}, narrow-band filtering \cite{siddons}, and stabilization of laser frequency \cite{wasik}.

Implementation of polarization-analysis techniques such as the crossed-polarizer (``forward scattering") configuration, balanced polarimeter and circular analyzer allows studying magnetic resonance and spectroscopic features, which carry information about the modification of dispersive (birefringence) and absorptive (dichroism) properties of atomic media, and permits to retrieve the Stokes parameters characterizing the state of light polarization \cite{huard}.

Most studies of magneto-optical effects in atomic vapor deal with longitudinal orientation of a magnetic field with respect to light propagation ($\textbf{B} \parallel \textbf{k}$), exploiting Faraday \cite{budker1} or Hanle \cite{dancheva} configurations, which yield narrow and high-contrast magnetic resonances around $B=0$. The appearance of sub-natural-width ($\sim$ 30 mG) magnetic resonances in the excited-state population for the case of linearly-polarized excitation of a transition with $F_e>F_g$ (where $F_{e,g}$ refer to the total angular momenta of the excited and the ground state, respectively) was described theoretically in \cite{renzoni}. It is caused by ground-state coherence effects, and was recorded experimentally in the probe-transmission spectra in \cite{grewal} when an external transverse magnetic field $\textbf{B} \perp \textbf{E}$ was applied.

A change of the light-induced atomic polarization occurs for an arbitrary direction of the external magnetic field \textbf{B}, except for the case where laser field \textbf{E} with wave vector \textbf{k} is polarized parallel to the transverse magnetic field ($\textbf{E} \parallel \textbf{B}$, $\textbf{B} \perp \textbf{k}$) \cite{nienhuis1}. The absence of Larmor precession of polarized atoms around the magnetic field in this case results in the absence of any non-spontaneous signal associated with magneto-optical rotation or induced ellipticity of resonant linearly polarized light. The latter seems to eliminate any possibility to generate narrow lines and thus makes such a field configuration unattractive for spectroscopic studies.

Nevertheless, narrow magnetic resonances with high signal-to-noise ratio may arise due to coherence effects related to partial inversion of population. This can occur on transitions between magnetic sublevels belonging to the excited and ground-state levels under linearly polarized intense excitation of cycling transitions of an alkali-atom D$_2$ line ($F_e > F_g$) \cite{movsisyan,gazazyan}. 

\begin{figure}[h!]
	\centering
	\begin{center}
		\includegraphics[width=200pt]{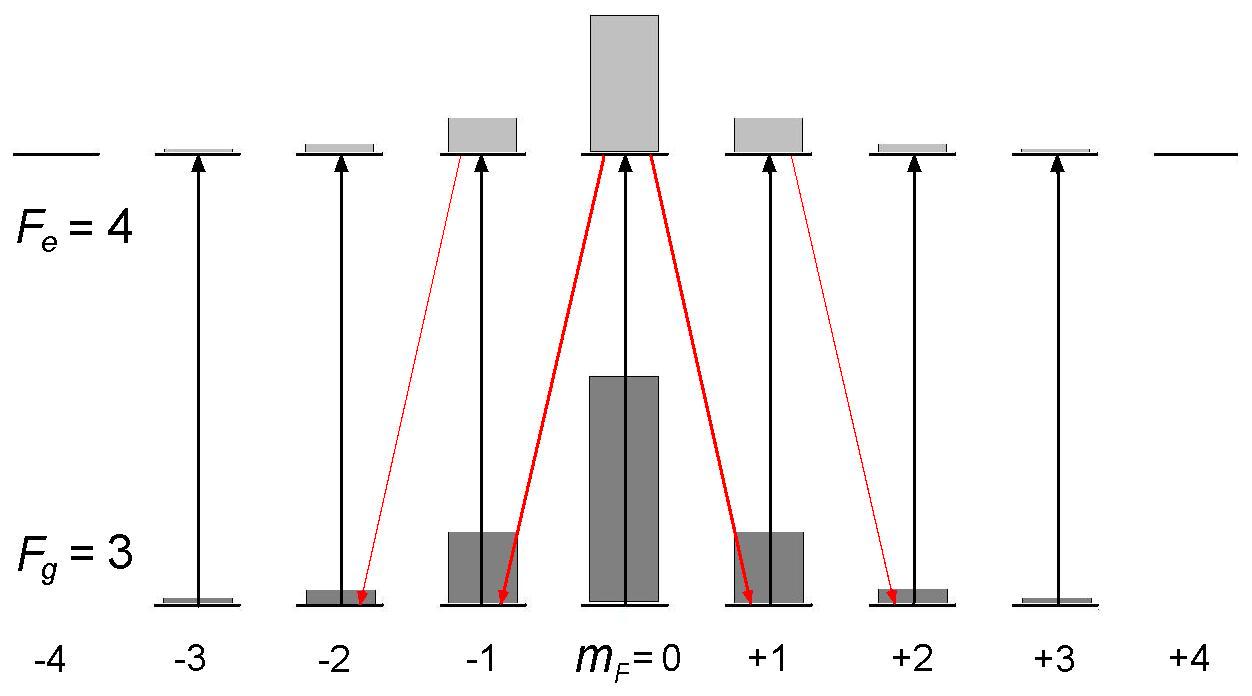}
		\caption{\label{fig:figure1} Onset of population inversion and amplified radiation corresponding to $\Delta m_F=\pm 1$ transitions under excitation of $^{85}$Rb D$_2$ line with an intense linearly polarized laser radiation for zero magnetic field (the quantization axis $Z$ is chosen to be along the laser light polarization).} 
	\end{center}
\end{figure}

In this paper, we report experimental observation of narrow (sub-natural), high-contrast magneto-optical resonances when irradiating rubidium vapor with an intense cw single-mode laser beam with a frequency tuned to $^{85}$Rb $F_g$=3 $\to$ $F_e$=4 or $^{87}$Rb $F_g$=2 $\to$ $F_e$=3 transitions of D$_{2}$ line. 
The magnitude and width of the resonances which are detected in a polarimetric scheme while scanning the transverse magnetic field (orthogonal to the laser beam propagation direction) around $B$ = 0, exhibit strong nonlinear dependence on the laser-light intensity and subnatural linewidth.

In \cite{movsisyan,gazazyan} it was demonstrated that interaction of linearly polarized light with a $F_e>F_g$ cycling D$_2$  transition of alkali atoms 
(in particular, $F_g$=3 $\to$ $F_e$=4 of $^{85}$Rb and $F_g$=2 $\to$ $F_e$=3 of $^{87}$Rb) may result in appearance of amplified radiation on transitions between certain Zeeman sublevels as shown in Fig.\,\ref{fig:figure1}. This amplification occurs at intensities of light above $\sim$ 20 mW/cm$^2$ in a steady-state regime due to redistribution of population towards $m_F$ = 0 \cite{taichenachev,nienhuis2}, which results in population inversion on $|m_{F_e}| = n \to |m_{F_g}| = n+1$, where $n$ is a non-negative integer.

\begin{figure}[h!]
	\centering
	\begin{center}
		\includegraphics[width=230pt]{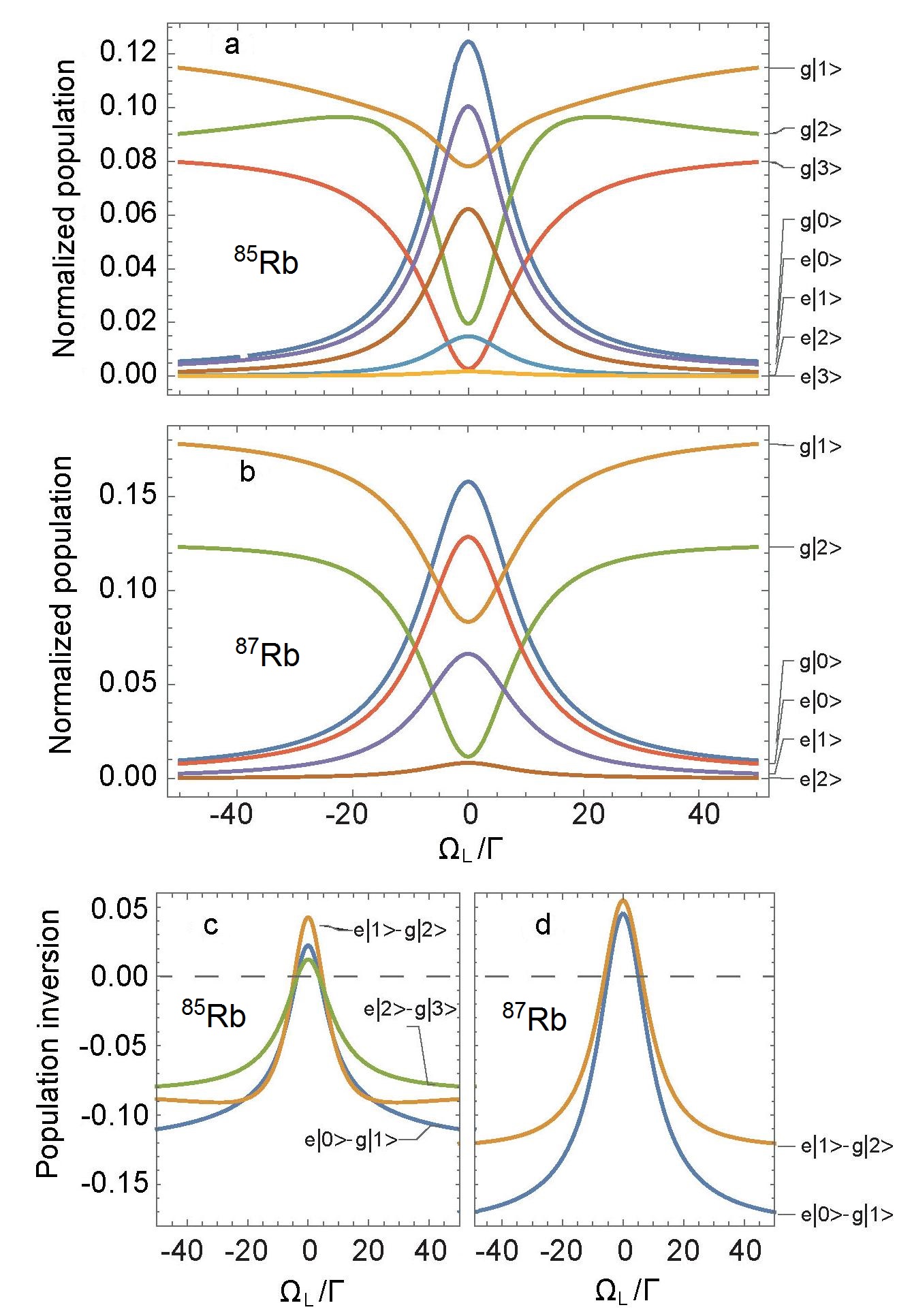}
		\caption{\label{fig:figure2}$\textbf{B} \parallel \textbf{E}$ magnetic-field dependence of normalized steady state population of ground state (g$|m_{F_g}$$>$) and excited state (e$|m_{F_e}$$>$) Zeeman sublevels linked by $\Delta m_F = 0$ transitions $^{85}$Rb $F_{g}$=3 $\rightarrow$ $F_e$=4 (a) and $^{87}$Rb $F_g$=2 $\rightarrow$ $F_e$=3 (b), and population inversion on corresponding $m_{F_e} \to m_{F_g}$ transitions (c,d). The light power is large enough to provide population inversion. $\Gamma$ is the decay rate of ground state, $\Omega_L$ is the Larmor frequency.}
	\end{center}
\end{figure}

The population inversion has strongly nonlinear dependence on the light intensity: above a threshold, the inversion increases reaching saturation followed by a decline. For $B=0$ and linearly polarized pump light, there appears optical gain for polarization orthogonal to that of the pump light in the forward and backward directions \cite{movsisyan}. Application of an external magnetic field should influence the redistribution of population due to Zeeman pumping in multiple absorption-emission cycles, thus affecting the onset of population inversion.

The case of magnetic-field-induced changes in transmitted light due to the gain that is most straightforward to interpret, involves a measurement configuration, which is least affected by concomitant magneto-optical effects. For the present study, we choose $\textbf{B} \parallel \textbf{E}$, where there is no optical rotation or ellipticity associated with the Faraday or Voigt effects \cite{nienhuis1}.

Theoretical consideration for the employed configuration confirms that magnetic field applied along the linear polarization of input light indeed affects establishment of population inversion. Dependences of the steady-state populations of particular Zeeman sublevels on $\textbf{B} \parallel \textbf{E}$ magnetic field under excitation with linearly-polarized light can be obtained by calculating the steady-state solution of the Liouville equation \cite{book}. The results of calculations using the Atomic Density Matrix (ADM) package \cite{simon} are shown in Fig. \ref{fig:figure2}. Magnetic field dependences of steady-state populations for the ground state sublevels ($m_{Fg}$) and excited state sublevels ($m_{Fe}$) of the ”cycling” transitions $F_{g}$=3 $\to$ $F_e$=4 ($^{85}$Rb) and $F_g$=2 $\to$ $F_e$=3 ($^{87}$Rb) of the rubidium D$_2$ line ($5S_{1/2} \to 5P_{3/2}$) under excitation with linearly-polarized radiation with $S=100$ are presented in Fig. \ref{fig:figure2}a and Fig. \ref{fig:figure2}b, respectively ($S$ is the saturation parameter defined as
\begin{equation}
S=\frac{|\Omega|^2}{\Gamma^2/4+\delta^2},
\end{equation} 
where $\Omega$ is the effective Rabi frequency, $\delta$ is the detuning from resonance including Doppler shift, and $\Gamma$ is the radiative decay rate).

The population inversion is shown in Fig. \ref{fig:figure2}c,d. As is seen from these graphs, the population inversion establishes at zero magnetic field, being stronger for the $^{87}$Rb $F_{g}$=2 $\to$ $F_e$=3 transition. Application of a magnetic field corresponding to $\Omega_L/\Gamma$ $\approx$ 5$\div$10 eliminates the population inversion.

Below, we present experimental data which indicate that, in this system, a number of different nonlinear-optical effects take place, including degenerate four-wave mixing and mirrorless lasing. The latter is a particularly interesting case because, in contrast to most other works on cw mirrorless lasing in alkali atoms (see \cite{akulshin} and references therein), in the present case, lasing occurs at the same wavelength as optical pumping, differing only in polarization (entirely degenerate mirrorless lasing).  

\section{Experimental arrangement}

\begin{figure}[h]
	\centering
	\begin{center}
		\includegraphics[width=250pt]{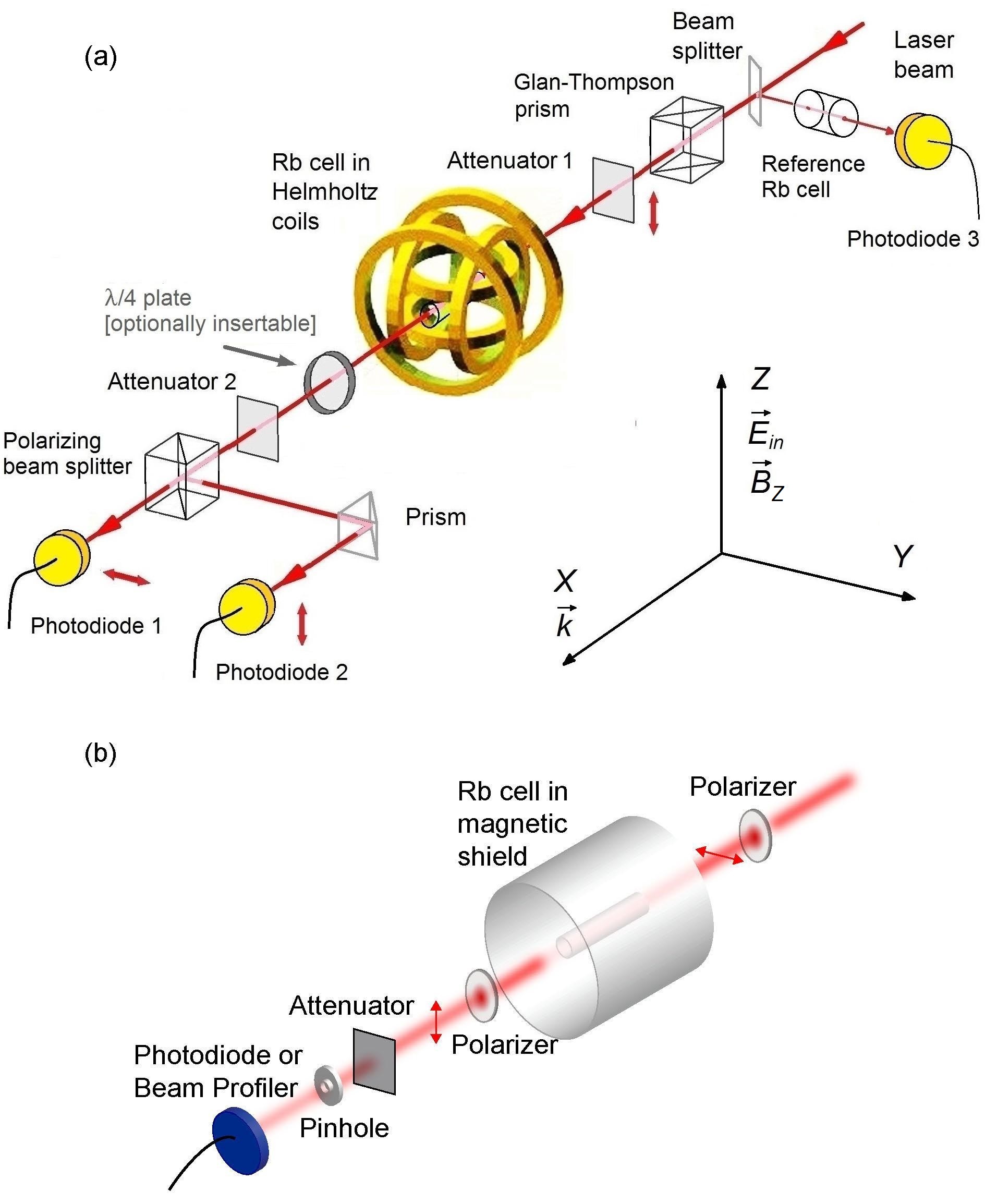}
		\caption{\label{fig:figure3} Experimental setup and the measurement configuration in Ashtarak (a) and Mainz (b).}
	\end{center}
\end{figure}

Experimental measurements were carried out on two independent setups, one in Ashtarak and another in Mainz, shown schematically in Fig. \ref{fig:figure3} (a and b, respectively). The results were initially obtained in Ashtarak, and were afterwards verified in Mainz, also exploiting additional capabilities such as better control of the applied magnetic field and precise measurements of the output light spatial profile.

In the Ashtarak setup (Fig. \ref{fig:figure3}a), a  $\approx$1.5-mm-diameter beam from a single-frequency cw extended-cavity diode laser (wavelength 780 nm, light power 20 mW, linewidth 1 MHz) was directed, after purification of the initial linear vertical polarization with a Glan-Thomson polarizer, onto a 135 mm-long sealed-off room-temperature glass cell with a side-arm containing natural rubidium. The cell had no coating or buffer gas, and was carefully checked to make sure that the mechanical stress-induced birefringence is negligible. 

The vapor-cell temperature was kept at 30$^{\circ}$C, which corresponds to a number density of $N$ = 1.16$\times10^{10}$ cm$^{-3}$ of Rb atoms. It was verified that small changes in atomic density did not qualitatively influence the observed effects. The cell was surrounded with an assembly of mutually orthogonal pairs of calibrated Helmholtz coils with the axes carefully aligned along the $X,Y,Z$ directions. The coils were used for canceling the magnetic field in two directions ($X,Y$), monitored to $\approx$ 10 mG with a magnetometer, and scanning within up to $\pm$ 1 G in the third direction ($Z$).

The light transmitted through the cell was directed onto a polarizing beam splitter decomposing the incident polarization into vertical and horizontal components that were simultaneously detected with two photodiodes. Diaphragms were placed in front of the photodiodes located 70 cm away from the cell in order to cut the contribution to the signal from the fluorescence of the Rb atoms. The two detection channels were adjusted to have the same sensitivity. 

A fraction of the laser light passed through an auxiliary room-temperature rubidium reference cell.

Mode-hop-free tuning range of the laser frequency covered the four spectrally resolved Doppler-overlapped transition groups of the Rb D$_2$ line: $^{87}$Rb $F_g$=2 $\to$ $F_e$=1,2,3; $^{85}$Rb $F_g$=3 $\to$ $F_e$=2,3,4; $^{85}$Rb $F_g$=2 $\to$ $F_e$=1,2,3; and $^{87}$Rb $F_g$=1 $\to$ $F_e$=0,1,2 (in order of increasing frequency). Each of these four lines is an overlap of three Doppler-broadened transitions among which one is cycling: $^{87}$Rb $F_g$=2 $\to$ $F_e$=3; $^{85}$Rb $F_g$=3 $\to$ $F_e$=4; $^{85}$Rb $F_g$=2 $\to$ $F_e$=1; and $^{87}$Rb $F_g$=1 $\to$ $F_e$=0. For sufficiently high laser intensity ($>$ 1 mW/cm$^{2}$), the non-cycling transitions essentially disappear due to depopulation optical pumping, so that the cycling transitions dominate the signals.
 
A set of calibrated neutral-density filters was used to change the incident light intensity (``Attenuator 1" in Fig. \ref{fig:figure3}a). The filters from the set, which were not used for attenuation, were placed at the position of ``Attenuator 2". With this configuration, the off-resonance signal did not change, so that there was no need to adjust the photodiode gain.

\begin{figure*}[ht]
	\centering
	\begin{center}
		\includegraphics[width=470pt]{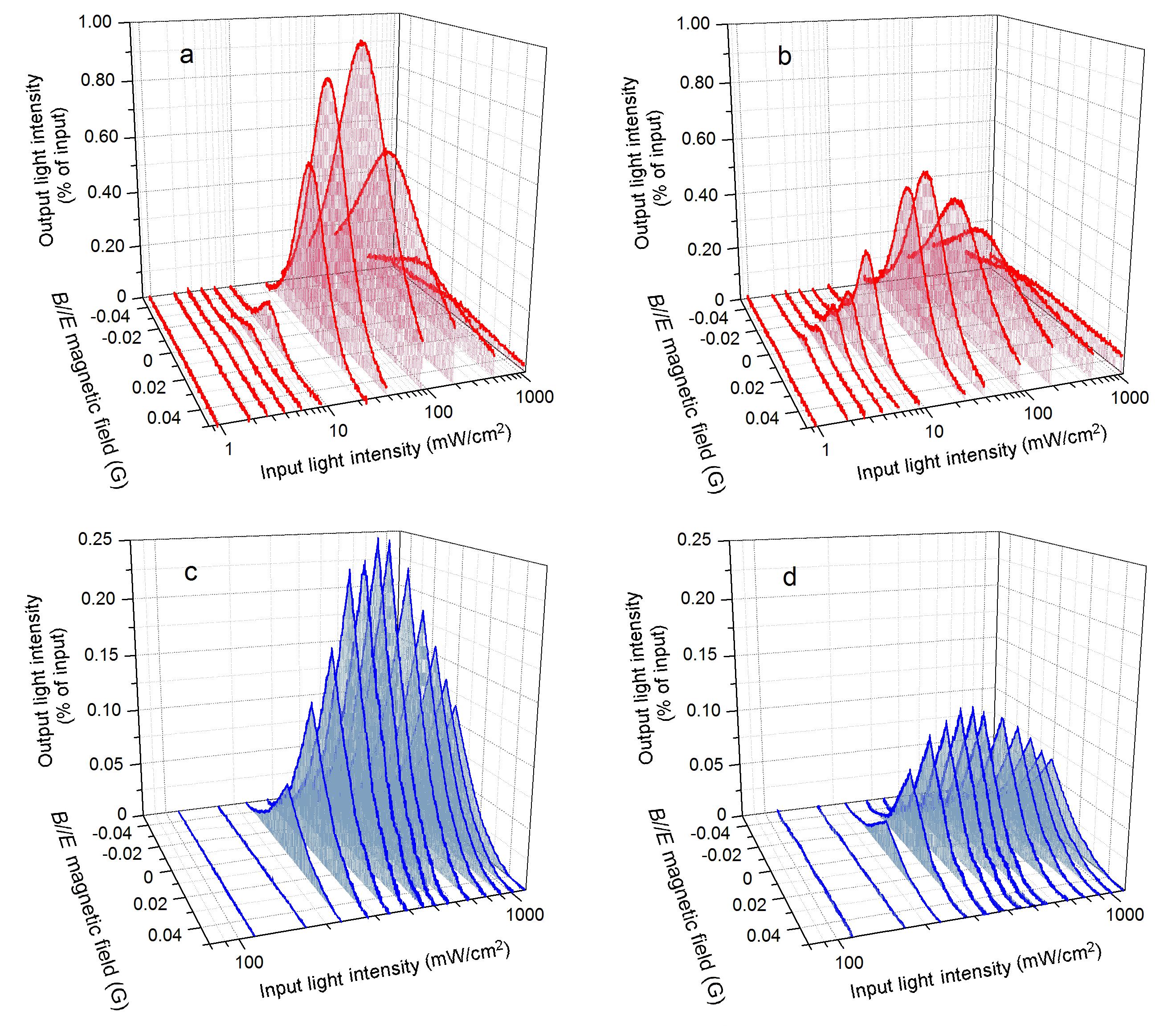}
		\caption{\label{fig:figure4} Dependences of normalized output intensity with orthogonal polarization (see text) on magnetic field ($\textbf{B} \parallel \textbf{E}$) for $^{85}$Rb $F_g$=3 $\to$ $F_e$=4 (a,c) and $^{87}$Rb $F_g$=2 $\to$ $F_e$=3 (b,d) D$_2$  transitions, measured in Ashtarak (a,b) and Mainz (c,d).}
	\end{center}
\end{figure*}

The Mainz setup (Fig. \ref{fig:figure3}b) was similar to that in Ashtarak, except for the magnetic shield. The mutually-orthogonal coils installed inside the shield allowed to apply and scan magnetic field in $X, Y, Z$ directions. Two polarizers (LPVIS050) were used for purification of the input laser light polarization and selection of the orthogonally-polarized component in the output beam. The spatial distribution of the forward beam emerging from the cell was studied with a beam profiler DMK 41BU02.H from Imaging Source. 
\section{Experimental results}

The first measurement on Ashtarak setup was carried out with scanning of $B_Z$ in the $\pm$ 0.12 G range. The frequency of the incident linearly polarized ($E_Z$) light was stabilized on the maximum of the Doppler-overlapped transition groups $^{85}$Rb $F_g$=3 $\to$ $F_e$=2,3,4 or $^{87}$Rb $F_g$=2 $\to$ $F_e$=1,2,3 transitions of Rb D$_2$ line, monitored with the reference signal. The intensity of the output light with orthogonal polarization ($E_Y$) was recorded with Photodiode 1 (see Fig. \ref{fig:figure3}a) for 12 values of incident laser intensity $I_{L}$ ranging from 1 to 1100 mW/cm$^2$.

\begin{figure}[h!]
	\centering
	\begin{center}
		\includegraphics[width=190pt]{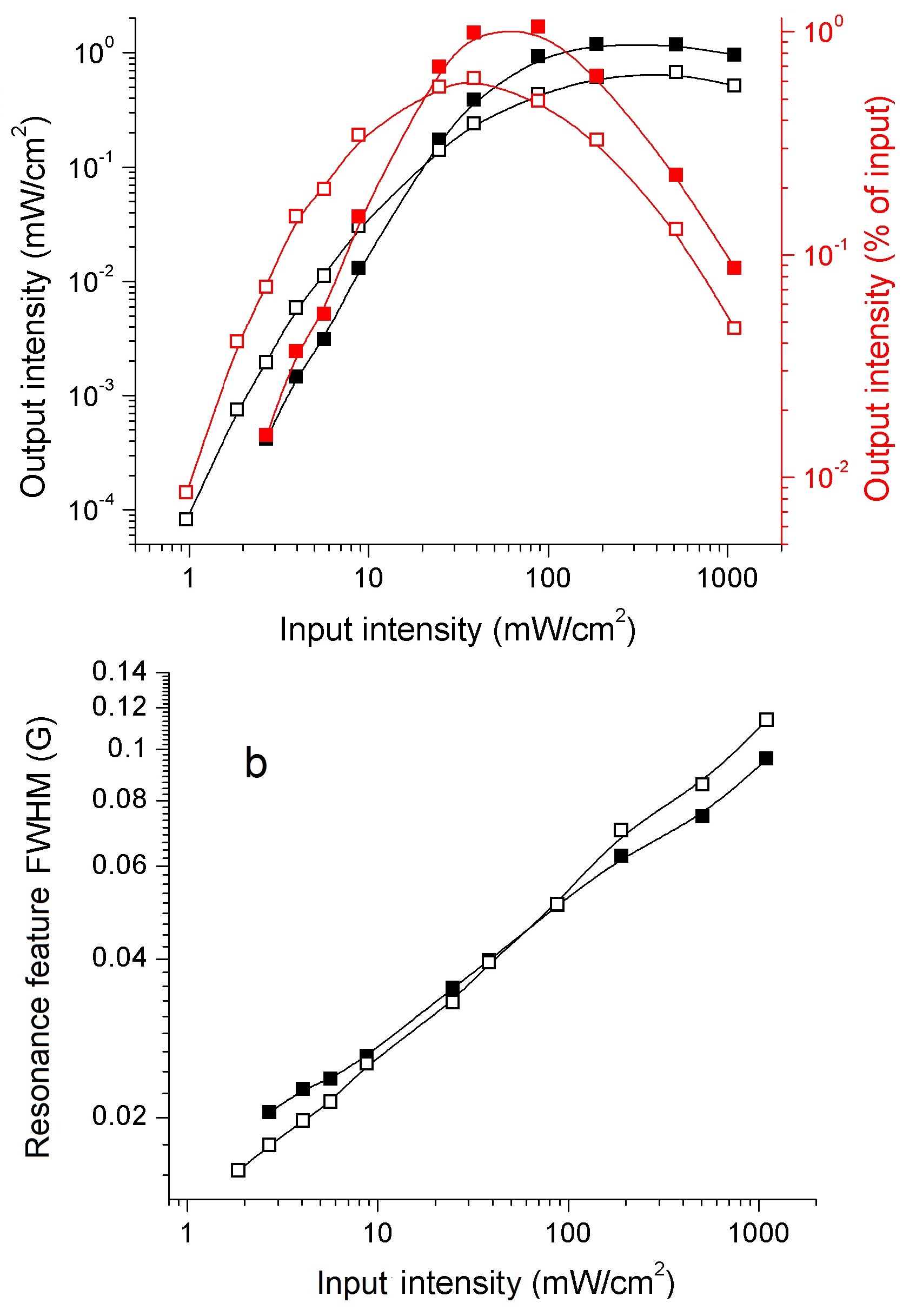}
		\caption{\label{fig:figure5} Input-intensity dependences of (a) absolute (black symbols, left axis) and normalized (red symbols, right axis) of the output intensity with orthogonal polarization for $B$ = 0; (b) FWHM width of $B$ = 0 resonance for the output orthogonally-polarized light, retrieved from the measurements presented in Fig. \ref{fig:figure4}a,b. Solid squares: $^{85}$Rb $F_g$=3 $\to$ $F_e$=4; open squares: $^{87}$Rb $F_g$=2 $\to$ $F_e$=3. Solid lines are drawn to guide the eye.}
	\end{center}
\end{figure}

The output intensity normalized by the input intensity is presented in Fig. \ref{fig:figure4}a,b. Negligible output signal was recorded for low laser radiation intensity ($I_{L}$ $<$ 1 mW/cm$^2$). At $I_{L}$ $\approx$ 1.5 mW/cm$^2$ a ($\approx$ 15 mG FWHM) resonance feature appears first on the $^{87}$Rb $F_g$=2 $\to$ $F_e$=3 transition, centered around $B$ = 0, and then also on the $^{85}$Rb $F_g$=3 $\to$ $F_e$=4 transition ($\approx$ 20 mG FWHM at $I_{L}$ $\approx$ 3 mW/cm$^2$). Further increase of intensity results in broadening and amplitude increase of the resonance feature to a maximum normalized value recorded at $I_{L}$ $\approx$ 35 and 85 mW/cm$^2$ for $^{87}$Rb $F_g$=2 $\to$ $F_e$=3 and $^{85}$Rb $F_g$=3 $\to$ $F_e$=4, respectively, followed by subsequent reduction of the transmitted intensity at $B$ = 0 and further broadening of the resonance.

Similar measurements carried out on the Mainz setup with the same cell length, temperature, and input beam diameter exhibit qualitatively similar behavior, with somewhat higher threshold and maximum conversion values of input light intensity (Fig. \ref{fig:figure4}c,d). Among other quantitative distinctions from Ashtarak experiment are lower maximum conversion efficiency ($\approx$ 0.25$\%$), and sharper shape of the $B$ = 0 peak. 

The scaled and absolute intensity values of $B$ = 0 resonances, as well as their widths depending on laser intensity, derived from the Ashtarak measurements presented in Fig. \ref{fig:figure4}a,b are shown in Fig. \ref{fig:figure5}. As seen in the figure, the absolute output intensity reaches 10 mW/cm$^2$ (or 1 $\%$ of the incident-light intensity). The input-intensity dependence of the output signal is strongly nonlinear (note the log-log scale). As to the resonance linewidth, it rises monotonically  from $\approx$ 0.02 to 0.1 G when changing $I_L$ by three orders of magnitude, from $\approx$ 1 to 1000 mW/cm$^2$, with a somewhat steeper slope for $^{87}$Rb $F_g$=2 $\to$ $F_e$=3.

Measurements similar to those presented in Fig. \ref{fig:figure4} were performed also for the output light with initial laser polarization ($E_Z$), recorded with Photodiode 2 (see Fig. \ref{fig:figure3}a). As expected, a dip centered at $B$ = 0 complementary to the peak for the orthogonal polarization was recorded for the same conditions and with nearly the same parameters (amplitude and width) as for the case of the orthogonal polarization. Unlike zero-background peaks for $E_Y$, the $E_Z$ dip appears on top of a background resulting in a low-contrast signal (not shown here).

\begin{figure}[h!]
	\centering
	\begin{center}
		\includegraphics[width=220pt]{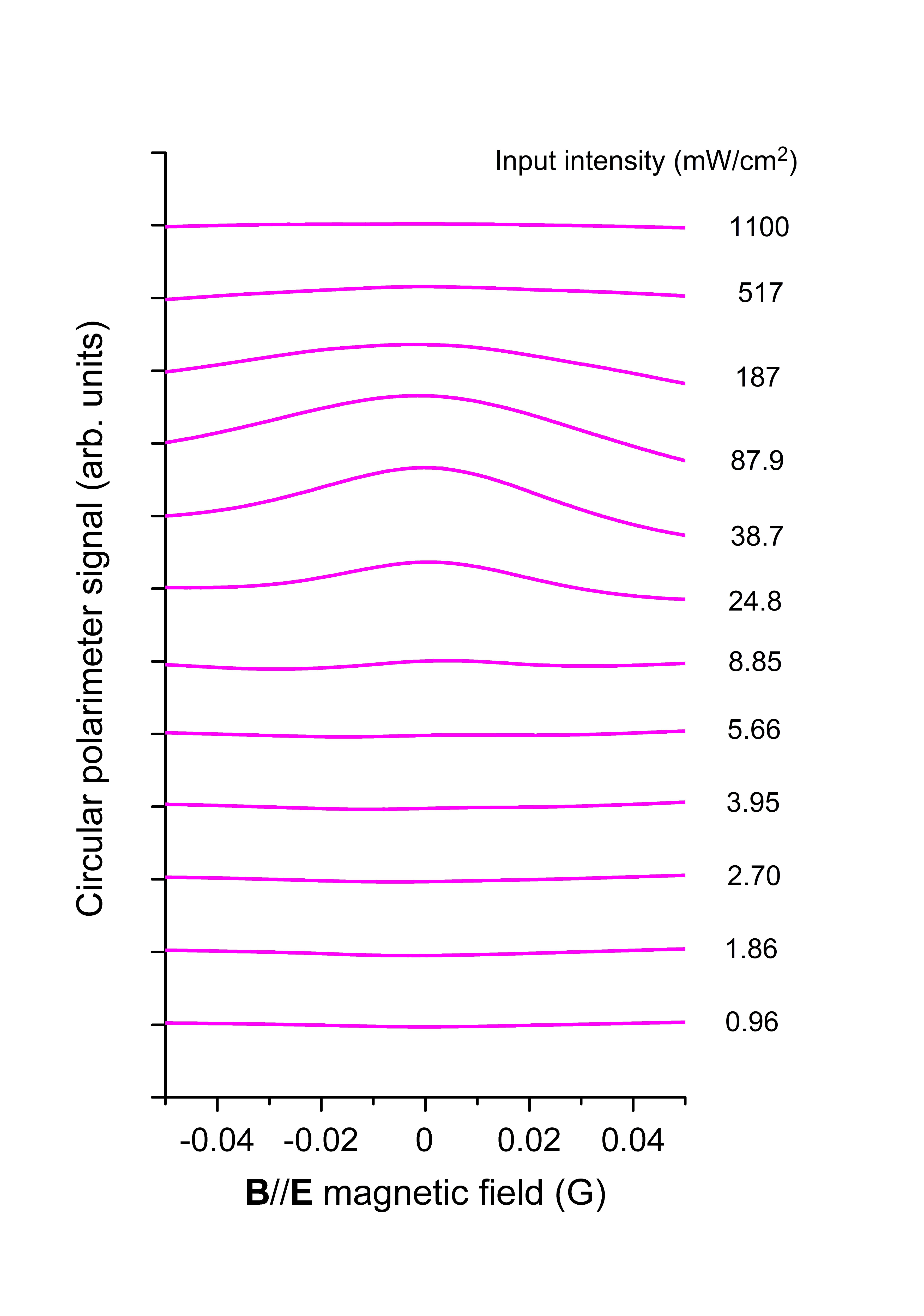}
		\caption{\label{fig:figure6} Magnetic-field dependence of the output signal for the $^{85}$Rb $F_g$=3 $\to$ $F_e$=4 transition recorded with a circular polarimeter at different input-light intensities (same as in Fig. \ref{fig:figure4}a,b). No normalization; the curves are shifted vertically by equal distances.}
	\end{center}
\end{figure}

Circular polarization analysis of the output light was also performed with the Ashtarak setup. Figure \ref{fig:figure6} presents the circular polarimeter output signal versus $\textbf{B} \parallel \textbf{E}$ for the case of $^{85}$Rb $F_g$=3 $\to$ $F_e$=4 transition measured at different values of $I_L$ corresponding to the conditions of Fig.\,\ref{fig:figure4}a,b. One can clearly see that ellipticity in output light polarization appears around $B$ = 0 for $I_L \approx$ 10 $-$ 500 mW/cm$^2$, i.e. in the input-intensity range where the orthogonally polarized output light is significant. This may be a signature of non-random phase of the generated light with respect to the excitation-light phase, and requires further experimental investigation.

\begin{figure*}[ht!]
	\centering
	\begin{center}
		\includegraphics[width=370pt]{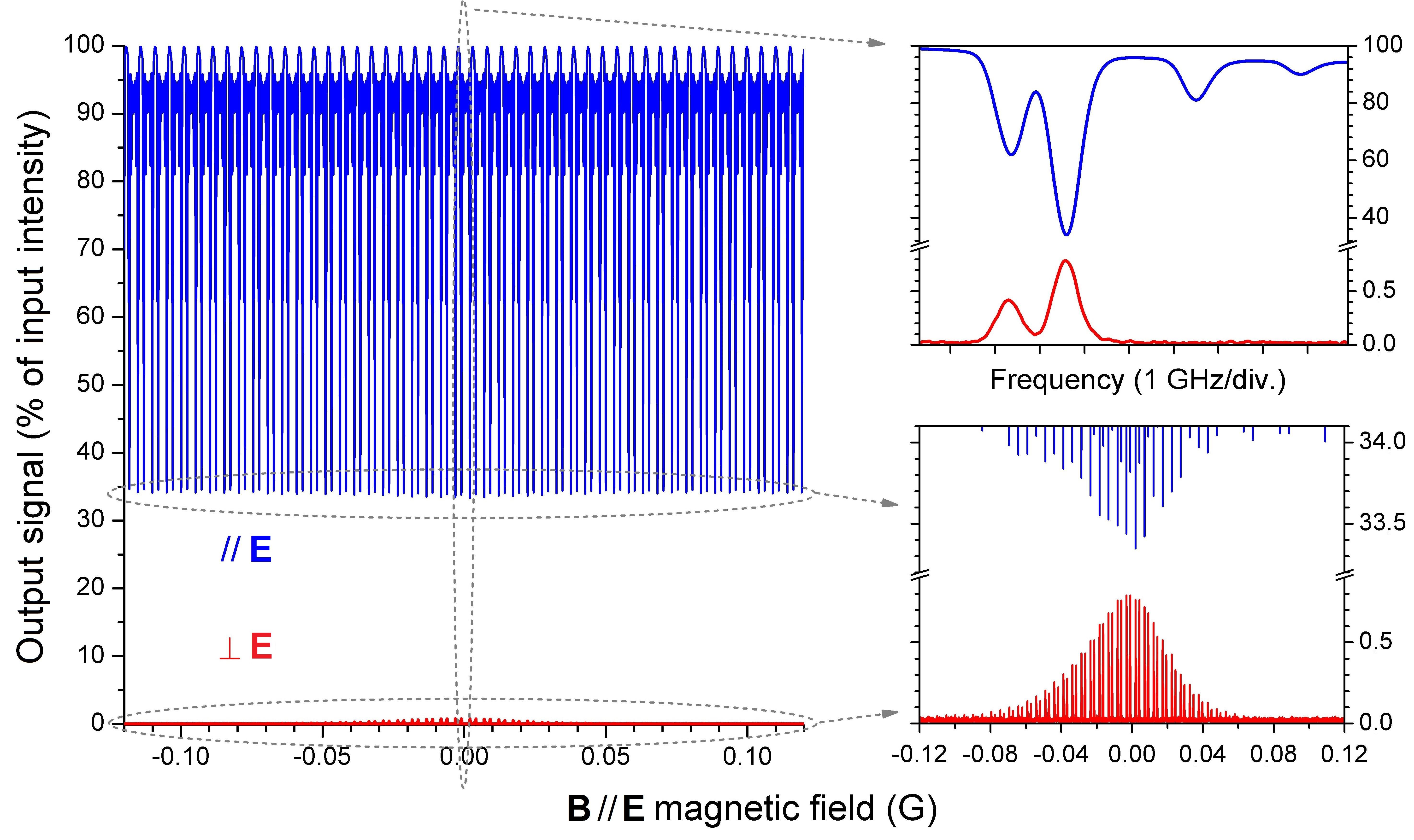}
		\caption{\label{fig:figure7} Double-scanning measurement for the output light with polarization parallel ($\parallel$$\textbf{E}$) and orthogonal ($\perp$$\textbf{E}$) to the input laser polarization for incident light intensity of 87.9 mW/cm$^2$. A single slow (compared to the frequency of the spectral scanning) $B$-field scan ($\textbf{B} \parallel \textbf{E}$) in the range of $\pm$0.12 G  is combined with fast laser-frequency scans over 10 GHz covering the Rb D$_2$ line. Upper inset: zoomed output light spectrum in the region of $B$ = 0; lower inset: zoomed signals of $\perp$$\textbf{E}$ output light (lower trace), and $\parallel$$\textbf{E}$ output light in the region of the dip (upper trace).}
	\end{center}
\end{figure*}

\begin{figure*}[ht!]
	\centering
	\begin{center}
		\includegraphics[width=370pt]{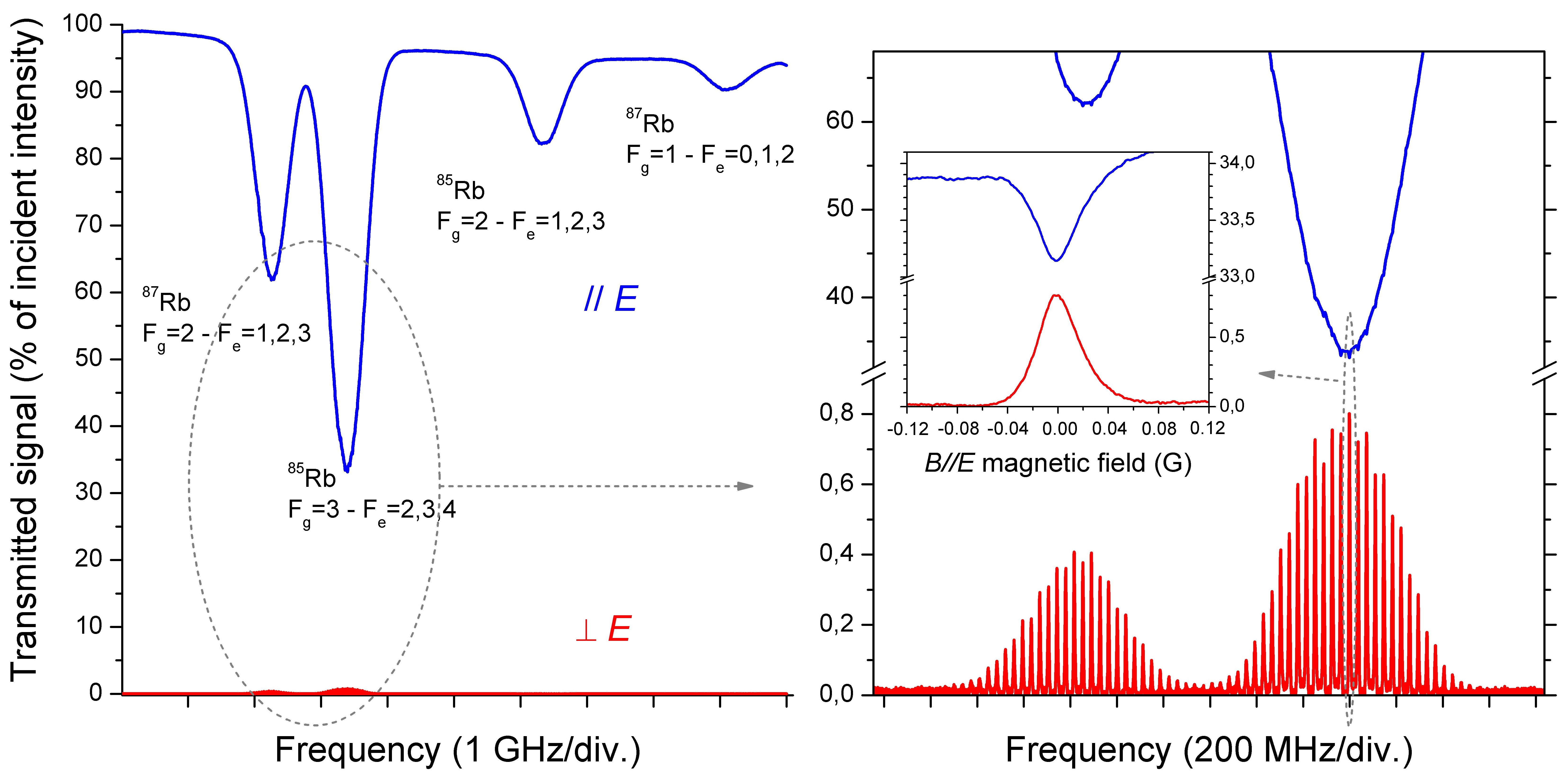}
		\caption{\label{fig:figure8} Left panel: double-scanning measurement for the output light with polarization parallel ($\parallel$$\textbf{E}$) and orthogonal ($\perp$$\textbf{E}$) to the input laser polarization for incident light intensity of 87.9 mW/cm$^2$. A single slow frequency scan with a 10-GHz span covering the D$_2$ line is combined with 270 fast $\pm$0.12 G $B$-field scans ($\textbf{B} \parallel \textbf{E}$). Right panel: zoomed-in spectra of $\perp$$\textbf{E}$ output light (lower trace), and $\parallel$$\textbf{E}$ output light in the region of the transmission dip (upper trace) covering the $^{87}$Rb $F_g$=2 $\to$ $F_e$=1,2,3 and $^{85}$Rb $F_g$=3 $\to$ $F_e$=2,3,4 transitions; inset: zoomed-in $^{85}$Rb $F_g$=3 $\to$ $F_e$=2,3,4 output signals.}
	\end{center}
\end{figure*}

The next series of measurements carried out in Ashtarak were aimed at detailed studies of the frequency dependence of the resonances across the whole spectral range of the Rb D$_2$ line. Simultaneous double linear scanning of laser frequency and $\textbf{B} \parallel \textbf{E}$ magnetic field was performed by applying triangular control pulses at different temporal rates.

\begin{figure*}[ht!]
	\centering
	\begin{center}
		\includegraphics[width=350pt]{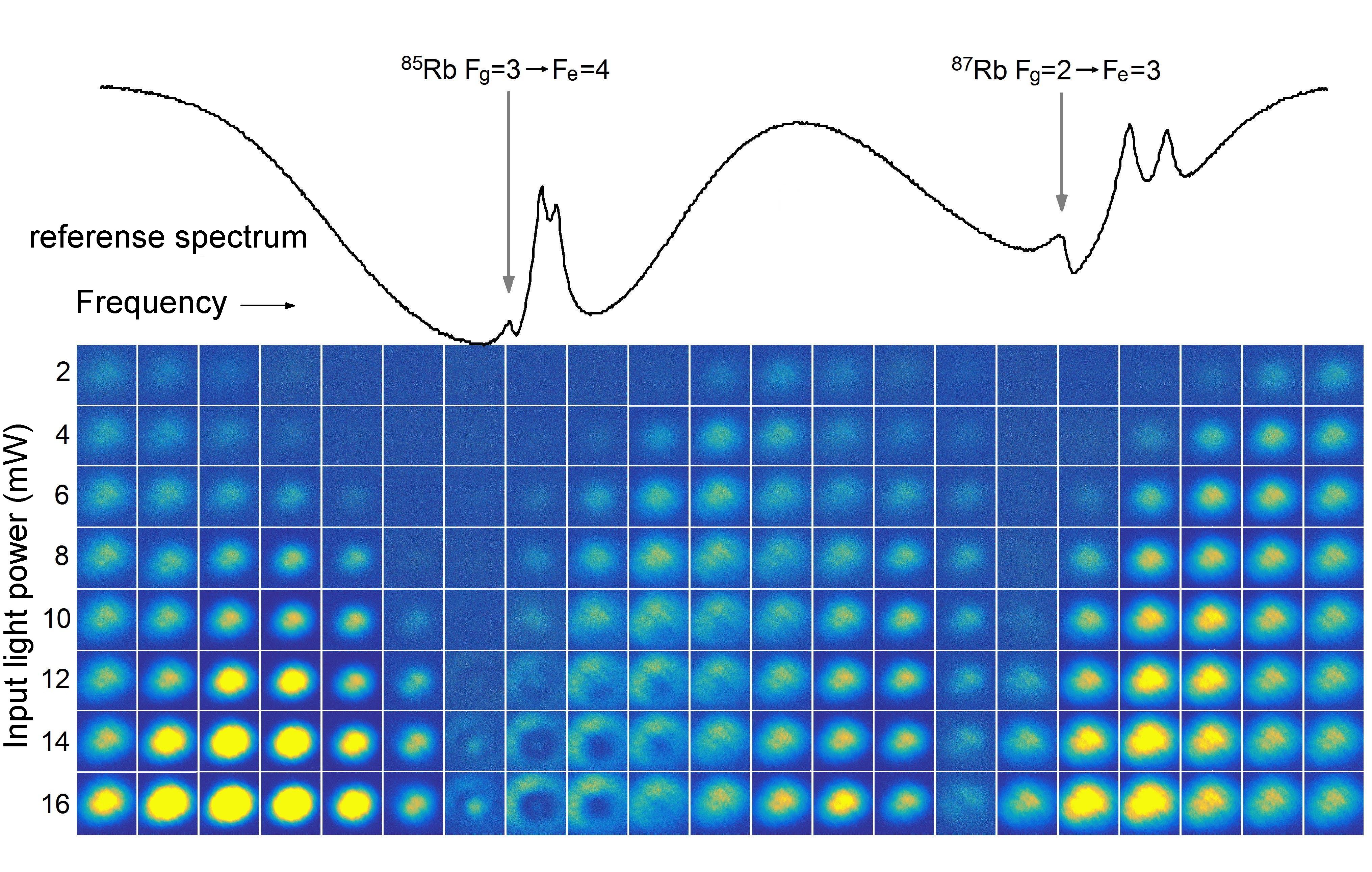}
		\caption{\label{fig:figure9} Spatial-profile measurements for zero-magnetic-field orthogonally polarized forward output beam versus the laser radiation frequency for different values of input light power. Upper curve: reference saturated-absorption spectrum.}
	\end{center}
\end{figure*}

\begin{figure*}[ht!]
	\centering
	\begin{center}
		\includegraphics[width=340pt]{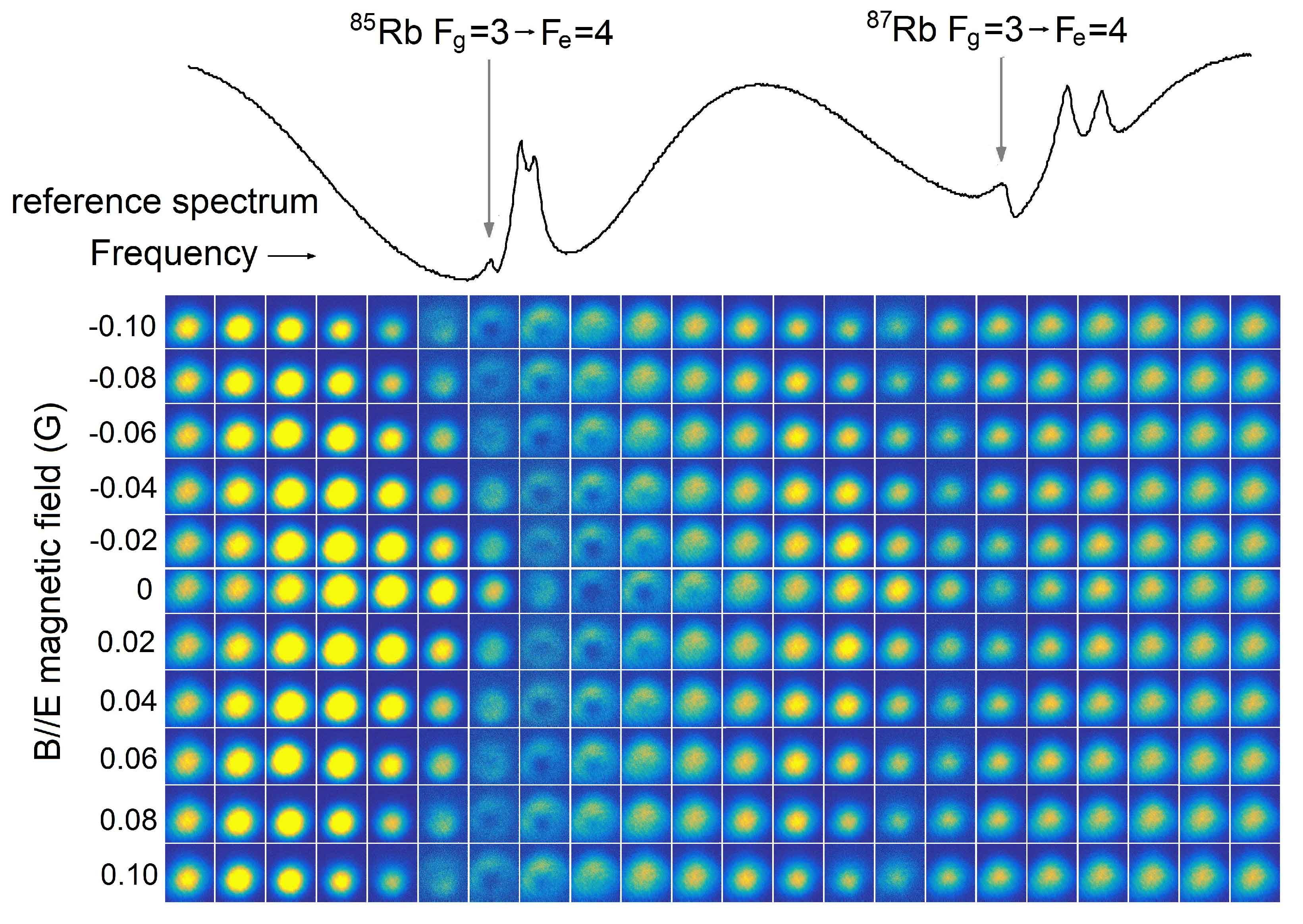}
		\caption{\label{fig:figure10} Spatial-profile measurements for the orthogonally polarized forward output beam versus the laser radiation frequency for different values of the $\textbf{B} \parallel \textbf{E}$ magnetic field, measured at 12 mW input light power. Upper curve: reference saturated-absorption spectrum.}
	\end{center}
\end{figure*}

The results of such measurements for the linear polarization components of the output light with simultaneous slow scanning of the $B$-field ($\textbf{B} \parallel \textbf{E}$) and fast scanning of the laser frequency across the D$_2$ line are presented in Fig. \ref{fig:figure7} for the incident intensity corresponding to maximum output signal with orthogonal polarization for the $^{85}$Rb $F_g$=3 $\to$ $F_e$=4 transition. It is seen from the upper zoomed-in inset that the $\perp\textbf{E}$ polarized-light spectrum (lower trace) does not contain the transition groups $^{85}$Rb $F_g$=2 $\to$ $F_e$=1,2,3 and $^{87}$Rb $F_g$=1 $\to$ $F_e$=0,1,2 unlike the $\parallel$$\textbf{E}$ polarized-light spectrum (upper trace), where these transitions are well pronounced. 

This peculiarity is also seen in Fig. \ref{fig:figure8} recorded for the same conditions with a single slow scan of laser frequency and continuously repeating fast $B$-field scans. Note that the maximal-amplitude of the resonance is observed when the laser frequency is tuned to the absorption maximum of the the cycling transitions. It is noteworthy that for the $B$ = 0 peak in the orthogonally-polarized output light there is a complementary dip of the same amplitude in signal with unchanged input polarization, as is clearly seen from the inset in the right panel of Fig. \ref{fig:figure8}. This is a direct signature of parallel-to-orthogonal polarization conversion in the output light. 

As mentioned above, there was a quantitative discrepancy between the Ashtarak and Mainz results for the input beam intensity values where the orthogonally-polarized output beam is efficiently generated, as well as for the conversion efficiency (Fig. \ref{fig:figure4}). One could ascribe this to the different output-beam detection geometries used on the two setups. The detection angle in Ashtarak was about 15 mrad, while in Mainz it was 6 mrad. Another possible reason is the magnetic-field environment (different precision of the $B$-field control). In order to clarify this issue, additional measurements were done in Mainz using a beam profiler for spatial analysis of the output beam.

The spatial profile of the orthogonally polarized output beam versus frequency, covering the $^{85}$Rb $F_g$=3 $\to$ $F_e$=2,3,4 and $^{87}$Rb $F_g$=2 $\to$ $F_e$=1,2,3 transitions groups is shown in Fig. \ref{fig:figure9} for different values of input beam power (no magnetic field applied). Across these measurements, the extinction ratio for the crossed polarizers was somewhat worsened by a slight tilt of the analyzer, so that when the laser radiation frequency is off-resonance from atomic transition, some residual light remains in the transmitted beam, with intensity increasing linearly with input power. When the laser frequency is tuned closer to the atomic transition, the residual light gets absorbed, but simultaneously a bright beam emerges above the threshold input power of $\approx$ 10 mW. In the region of $^{85}$Rb $F_g$=3 $\to$ $F_e$=4 transition, the emerging narrow axial beam is surrounded by a ring structure, which remains, even becoming stronger, with some blue frequency shift from the resonance. The threshold of the ring structure is somewhat different from the one of the central beam. The ring structure is not observed in the region of $^{87}$Rb $F_g$=2 $\to$ $F_e$=3 transition.

The same double-scanning technique was employed for the study of the output beam profile versus radiation frequency and $\textbf{B} \parallel \textbf{E}$ magnetic field. For this measurement, the value of input light power was kept unchanged at 12 mW (above the threshold for the $^{85}$Rb $F_g$=3 $\to$ $F_e$=4 transition). The results of this measurement are shown in  Fig. \ref{fig:figure10}. One can clearly see that the increase of the magnetic field leads to vanishing of the central beam, unlike the ring structure which becomes even more pronounced.

The results of these measurements show that detection angle (aperture) is indeed of great importance for the studied process, which may explain the quantitative distinction of Ashtarak and Mainz results observed in Fig. \ref{fig:figure4}.

\section{Discussion of the results}

We believe the obtained experimental results are consistent with the mechanism described in the Introduction, sufficiently proving the link between the observed features with the formation of amplified orthogonally polarized light (mirrorless lasing) under excitation of an $F_e > F_g$ atomic cycling system with intense linearly polarized light.

Linearly polarized light induces polarization of the $F_e > F_g$ atomic system resulting in light-induced alignment, redistributing the ground state population towards the $m_{F_g}$ = 0 Zeeman sublevel. In the weak-light-intensity limit, application of weak $B_Z$ ($\textbf{B} \parallel \textbf{E}$) magnetic field does not influence transmitted radiation because of symmetry. When the excitation intensity is strong enough (above saturation), light is generated, which is orthogonally polarized with respect to the incident radiation. This results in an atomic polarization component that can undergo Larmor precession around $B_Z$. The latter tends to redistribute the ground-state population among Zeeman sublevels, eventually affecting conditions of population inversion. The characteristic strength of magnetic field $B_c$ needed to destroy amplification is determined by the ground-state relaxation rate 
(time-of-flight broadening for an atom traversing the laser beam $\gamma_c/2\pi \approx$ 100 kHz).
The magnetic-resonance width of $B_c\approx$ 20\,mG estimated from the ground-state relaxation rate due to the atoms' transit across the laser beam is in agreement with the experimentally observed FWHM widths presented in Fig. \ref{fig:figure5}b at the low-intensity limit.


Contribution of this process in our results is evidenced by sharp $B$ = 0 peaks in Fig. \ref{fig:figure4}. Blurry peak features observed in the Ashtarak experiment (Fig. \ref{fig:figure4} a,b) are caused by the absence of shielding, which results in a residual ac components of laboratory magnetic field estimated to be at the 10 mG level. Note that the coherence time can be substantially increased (by orders of magnitude) using antirelaxation-coated cells, cells with buffer gas, or cold atoms \cite{budker1}, allowing reduction of magnetic resonance width down to $\sim$ 0.1 mG.

There is yet another mechanism responsible for effective broadening of the magnetic resonance with $I_L$. At high light intensity, the distribution of the ground-state Zeeman sublevel populations is dominated by light-induced polarization, thus diminishing the effect of Larmor coupling. As a result, the alignment remains preserved for larger range of the $B$-field. This causes continuous growth of the magnetic-resonance width as a function of light power (Fig. \ref{fig:figure5}b), as opposed to the output light intensity, which undergoes saturation at high values of $I_L$ (Fig. \ref{fig:figure5}a).  

Comparison of Fig. \ref{fig:figure4} a and b shows that the $B$ = 0 peak signal appears at lower values of $I_L$ for $^{87}$Rb $F_g$=2 $\to$ $F_e$=3 as compared with $^{85}$Rb $F_g$=3 $\to$ $F_e$=4. This result is in qualitative agreement with the calculations presented in \cite{movsisyan}, which give the values for inversion threshold intensity $S \approx$13 and $\approx$25, respectively. 

The suppression of orthogonally polarized amplified radiation at $B$ = 0 observed for $I_L$ $>$ 100 mW/cm$^2$ (Fig. \ref{fig:figure5}a) is in qualitative agreement with the experimental results obtained in \cite{movsisyan}. This decline may be attributed to saturation effects, and will be investigated in more detail in future work.

Generation of the orthogonally polarized output beam at $B$ = 0 should cause ellipticity in the overall polarization of the transmitted beam, which is consistent with the measurements done with circular polarimeter configuration (Fig. \ref{fig:figure6}): the ellipticity with up to $\approx$ 1:100 axis ratio is observed only in the conditions favorable for mirrorless lasing. Application of $B_Z$ magnetic field tends to cease the generation, thus recovering linear polarization of the output beam and giving rise to $B$ = 0 peak feature observable in the figure.

For non-cycling transitions ($^{85}$Rb $F_g$=2 $\to$ $F_e$=2, $F_g$=2 $\to$ $F_e$=1 and $F_g$=1 $\to$ $F_e$=1 and $^{87}$Rb $F_g$=1 $\to$ $F_e$=0 and $F_g$=1 $\to$ $F_e$=1) pumped with a linearly polarized laser light, practically no atoms remain in the excited state (see, for example, \cite{nienhuis2}). Consequently, no population inversion can develop. Indeed, no mirrorless lasing is observed on these transitions.

\begin{figure}[ht!]
	\centering
	\begin{center}
		\includegraphics[width=210pt]{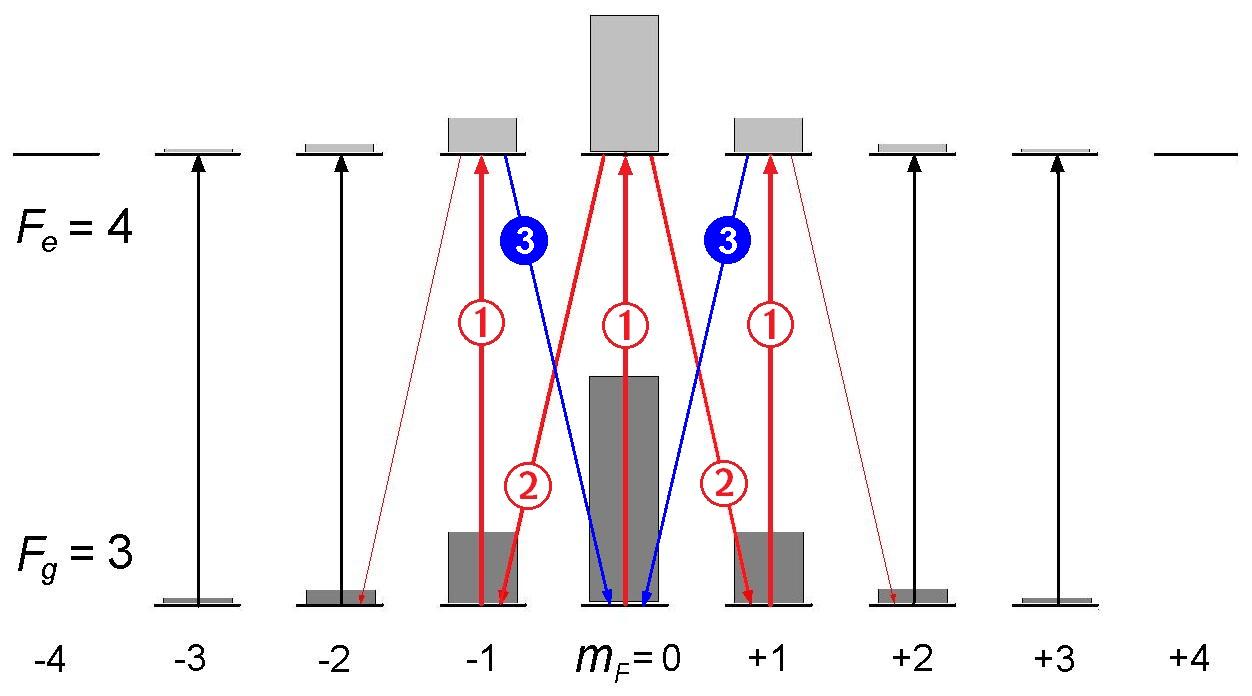}
		\caption{\label{fig:figure11} Schematic diagram of entirely-degenerate four-wave mixing process.}
	\end{center}
\end{figure}

Finally, the ring structure observed in the spatial profile of the output beam in the conditions close to the optimal conditions for mirrorless lasing is an evidence for a four-wave mixing (FWM) process which can develop in a system of magnetic sublevels of degenerate two-level atoms pumped with intense linearly polarized light. A schematic diagram of this process is presented in Fig. \ref{fig:figure11}. Excitation on transition $\textcircled{1}$ drives atoms from $m_{F_g}$ = 0 sublevel to $m_{F_e}$ = 0, followed by transfer to $m_{F_g}$ = $\pm$1 by inversion-amplified radiation $\textcircled{2}$ with subsequent laser excitation to $m_{F_e}$ = $\pm$1, and finally the loop is closing by emission of induced light $\textcircled{3}$. The figure shows only two symmetric FWM channels linked with $m_F$ = 0, but the process can similarly occur also on neighboring magnetic sublevels. We should note the distinction of this process from other known schemes: it is $entirely$ degenerate (all the four waves have the same frequency, they differ only by polarization).

Efficient generation of FWM requires phase-matching, the conditions for which are related to the dispersive properties of the resonant medium. Conical emission results from phase matching and can be straightforwardly separated from the axial mirrorless lasing, the latter having much lower divergence. Being assisted by the population inversion, the FWM process competes with the direct mirrorless lasing, which results in different optimal conditions for their observation as is seen in Figs.\,\ref{fig:figure9} and \ref{fig:figure10}. 

Summarizing, let us list the evidence supporting the conclusion that we have observed degenerate mirrorless lasing.

1) The absence of magnetic resonance for $B_Y$ ($\textbf{B} \parallel \textbf{E}$) below $I_{L}$ $\approx$ 1.5 mW/cm$^2$ ($^{87}$Rb $F_g$=2 $\to$ $F_e$=3) and $\approx$3 mW/cm$^2$ ($^{85}$Rb $F_g$=3 $\to$ $F_e$=4); nearly exponential initial above-threshold growth of the $B$ = 0 feature until saturation intensity (Figs. \ref{fig:figure4}, \ref{fig:figure5}a).

2) There is a narrow resonance in the magnetic-field dependence of the generated light in a configuration, where no such resonance is expected in this field range for non-lasing processes for symmetry reasons ($B$-field is directed along the light polarization).

3) Consistency of the laser-intensity range where magnetic resonance is observed (Fig. \ref{fig:figure4}) with the results of \cite{movsisyan}, where amplification due to partial population inversion on $^{85}$Rb $F_g$=3 $\to$ $F_e$=4 and $^{87}$Rb $F_g$=2 $\to$ $F_e$=3 transitions was directly recorded in the backward direction. 

4) Outright absence of any orthogonally polarized output light for $^{85}$Rb $F_g$=2 $\to$ $F_e$=1,2,3 and $^{87}$Rb $F_g$=1 $\to$ $F_e$=0,1,2\,($F_e < F_g$) transitions, independently of the $B_Z$ value (Figs.\,\ref{fig:figure7}, \ref{fig:figure8}).


A comprehensive theoretical treatment and modeling of the problem is complicated by complexity of the processes to be considered, including propagation effects with partially depleting pumping, and will be the subject of future work.

\section{Conclusions}

In conclusion, we have experimentally observed degenerate mirrorless forward lasing under excitation of rubidium vapor contained in uncoated buffer-gas-free vapor cells with linearly polarized laser radiation tuned to $^{85}$Rb $F_g$=3 $\to$ $F_e$=4 and $^{87}$Rb $F_g$=2 $\to$ $F_e$=3 D$_2$ transitions. The generated beam is orthogonally polarized to the incident light polarization, and has low divergence ($\approx$$10^{-3}$\,rad). The lasing occurs above certain threshold value of incident light intensity ($I_L \sim$ 3 mW/cm$^2$), and grows nonlinearly at least up to $\sim$ 0.3 $\%$ conversion efficiency at $I_L \approx$ 90 mW/cm$^2$. Further increase of $I_L$ results in a decrease of the intensity of the generated light, which almost completely vanishes at $I_L \sim$ 1000 mW/cm$^2$. The generated light is sensitive to application of a transverse ($\textbf{B} \parallel \textbf{E}$) magnetic field, forming  a narrow ($\sim$ 20 mG) resonance centered at $B$ = 0. 

Based on the analysis of the experimental results, the observed effect is attributed to the population inversion established on $|m_{F_e}| = n \to |m_{F_g}| = n+1$ transitions between the Zeeman sublevels ($n$ is non-negative integer), resulting in an orthogonally-polarized amplified beam under excitation of D$_2$ $F_e > F_g$ transition with linearly-polarized laser light.

The obtained results can be of interest for quantum-information applications (generation of entangled photon pairs and possibly dark resonances due to the presence of multiple coherent pathways of generated light). After further elaboration using antirelaxation-coated or buffered vapor cells the obtained results may be of interest for optical magnetometry.

Future work will include comprehensive studies of mirrorless lasing in the backward direction, which can be applied for remote mesospheric magnetic sensing using sodium layer as a sensor \cite{bustos}.

\textbf{Acknowledgments.} The authors are grateful to D. Sarkisyan, G. Grigoryan, and A.M. Akulshin for stimulating discussions, and to E. Klinger for assistance in experiment. Guzhi Bao acknowledges support by the German Federal Ministry of Education and Research (BMBF) within the “Quantumtechnologien” program (FKZ 13N14439).
This work was supported by the Committee of Science of Ministry of Education and Science of Armenia, and the German Federal Ministry of Education and Research, in the frame of the Armenian-German research project OMNIDAV (No. 15T-1C277, No. 16GE-035 /  01DK17057).

\end{document}